%% file: ghag.tex
\documentclass[11pt]{article}
\usepackage{xspace}
\usepackage{graphicx}
\usepackage{amsmath}
\usepackage{amssymb}
\usepackage{color}

\input econfmacros.tex
\def\Title#1{\begin{center} {\Large {\bf #1} } \end{center}}

\begin{document}

\Title{The LUX direct dark matter search experiment}

\bigskip\bigskip


\begin{raggedright}  

Chamkaur Ghag\index{Ghag, C.}, {\it University College London}\\

\begin{center}\emph{On the behalf of the LUX Collaboration.}\end{center}
\bigskip
\end{raggedright}

{\small
\begin{flushleft}
\emph{To appear in the proceedings of the Interplay between Particle and Astroparticle Physics workshop, 18 -- 22 August, 2014, held at Queen Mary University of London, UK.}
\end{flushleft}
}

\section{Introduction} 

The $\Lambda$-Cold Dark Matter standard model of Big Bang cosmology tells us that we live in an inflationary universe that is made up of contributions from dark energy (68\%), responsible for the accelerating expansion of the Universe, baryonic matter (5\%), and a dark matter component, which makes up the remaining 27\%. The evidence in support of the presence of dark matter is abundant and varied, and includes galactic rotation curves, measurements of the cosmic microwave background, weak lensing studies of galaxy clusters, primordial nucleosynthesis and the characteristics of large scale structure in the universe~\cite{Agashe:2014kda}. Despite considerable knowledge concerning the impact of dark matter on these astrophysical phenomena very little is known about its fundamental nature. Direct search experiments aim to detect individual interactions of particles of dark matter that are hypothesised to permeate our galaxy. Many experiments focus on the search for Weakly Interacting Massive Particles (WIMPs), the leading candidates for dark matter. They look for the low energy nuclear recoils expected when WIMPs scatter elastically off target nuclei in the experiment. The small interaction cross sections and low velocities expected for galactic WIMPs impose the challenging requirement that dark matter detectors need to be sensitive to $\sim$few keV recoiling nuclei and at the same time be capable of amassing exposures of many $\mathrm{kg} \cdot \mathrm{years}$. 

\section{The LUX Experiment} 

The Large Underground Xenon (LUX) experiment operates a dual-phase (liquid/gas) xenon time projection chamber (TPC) located 4850 feet underground (4300 m w.e.) at the Sanford Underground Research Facility (SURF) in Lead, South Dakota. The active region of the TPC is 47 cm in diameter and 48 cm in height comprising 250~kg of the 370~kg total xenon in the detector. Interactions in liquid xenon TPCs generate both prompt scintillation light (S1) and ionisation electrons that drift in an applied electric field (181 V/cm in LUX) to the liquid-gas interface at the top of the detector~\cite{Chepel:2012sj}. The electrons are then extracted into the gas phase (6.0 kV/cm in LUX), where they produce electroluminescence (S2). The S1 and S2 signals are used to reconstruct the deposited energy and their ratio is used to discriminate WIMP-like nuclear recoils (NR) from background electron recoils (ER). In LUX discrimination is achieved at the 99.6\% level for a 50\% NR acceptance in the energy range of interest. The TPC is viewed by two arrays of 61 photomultiplier tubes (PMTs) which image the central liquid xenon region from above and below and record the S1 and S2 signals. The \emph{x-y} position of an interaction is determined  from the localisation of the hit pattern of S2 light in the top PMT array. The depth of the interaction is measured through the drift speed of the electrons ($1.51 \pm 0.01$ mm/$\mu$s) and the time interval between the S1 and S2 light. This knowledge of the 3D position of an interaction, to about 4--6 mm in both \emph{x-y} and \emph{z}, means the powerful self-shielding capability of liquid xenon can be exploited to define an inner radioactively-quiet fiducial volume in which to perform the WIMP search. 

An extensive screening campaign imposed stringent requirements on the levels of radioactivity for materials used to build the detector. Before being used in LUX, the full contingent of research grade xenon was purified at a dedicated research facility using a novel technique based on chromatographic separation. In addition to shielding against cosmic rays provided by the rock overburden, the LUX detector sits within a 6.1 m tall and 7.6 m in diameter water tank, instrumented with 20 8-inch PMTs. This acts as both an active veto for any penetrating cosmic rays and as a shield against ambient $\gamma$-rays and neutrons. External backgrounds are thereby rendered subdominant to those from radioactivity within the xenon or from detector components. A full description of LUX can be found in~\cite{Akerib:2012ys}.

\section{First results from LUX} 

LUX completed its first physics run in 2013, collecting a total of 85.3 live-days of WIMP search data between late April and early August. During this period the ER background rate inside the 118 kg fiducial volume was measured to be $3.6\pm0.3$ mDRU (mDRU$=10^{-3}$ counts/day/kg/keV) in the energy range of interest, to date the lowest achieved by any xenon TPC. Full details of the radiogenic and muon-induced backgrounds in LUX can be found in~\cite{Akerib:2014rda}. A non-blind analysis was conducted in which only a minimal set of high-acceptance data quality cuts were used. Single scatter events containing exactly one S1 within the maximum drift time (324 $\mu$s) preceding a single S2 were selected for further analysis. The single scatter ER and NR acceptance was measured with dedicated tritium ($\beta^{-}$), AmBe, and $^{252}$Cf (neutron) datasets. All the cuts and efficiencies combined to give an overall WIMP-detection efficiency of 17, 50 and $>$ 95\% at 3.0, 4.3 and 7.5 $\mathrm{keV}$ recoil energies, respectively.

\begin{figure}[htp!]
\centerline{
\includegraphics[trim=0 2 0 6,clip, width=0.9\textwidth]{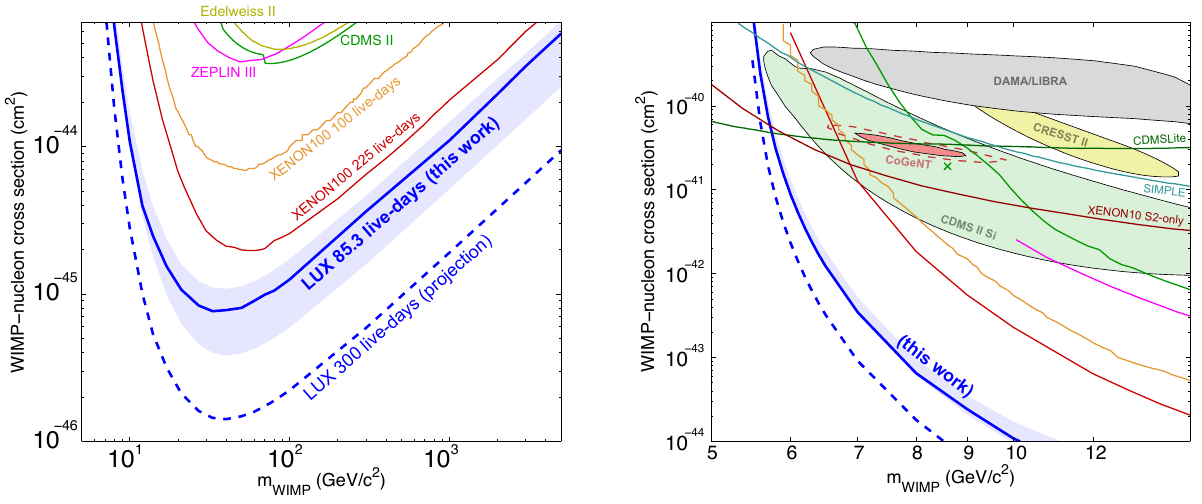}
}
\vspace{-10pt}
\caption{{\em Left:} The LUX 90\% confidence limit on the spin-independent elastic WIMP-nucleon cross section for the 85.3 live-day exposure (blue) and projected limit for the upcoming 300-day run (dashed blue). {\em Right:} Zoom of the low-mass region.}
\label{fig:limits}
\end{figure}

A total of 160 events were observed between 2--30 photoelectrons (phe) S1, the energy range of interest for WIMPs, with the observed rate and distribution found to be consistent with the predicted background of electron recoils. The p-value for the background-only hypothesis was 0.35. Confidence intervals on the spin-independent WIMP-nucleon cross section were set using a profile likelihood ratio (PLR) test statistic which exploits the separation of signal and background distributions in radius, depth and S1 and S2. For the signal model we conservatively assumed no signal below 3 keV, the lowest energy for which direct light yield measurements in xenon existed at that time. The 90\% upper C.L. are shown in figure~\ref{fig:limits} (left) with a minimum of $7.6 \times 10^{-46} \mathrm{cm}^{2}$ at a WIMP mass of 33 GeV$/\mathrm{c}^2$, the world's most stringent direct constraint on WIMP-nucleon interactions to-date. Figure~\ref{fig:limits} (right) shows the LUX constraints at low WIMP masses, excluding the majority of parameter space postulated as possible signal claimed by a number of experiments. Full details of the analysis can be found in~\cite{Akerib:2013tjd}. 

Following the first WIMP-search result LUX underwent a period of calibration data taking, maintenance, and optimisation of operational parameters in preparation for its primary 300-day WIMP search that runs into 2015. This included a campaign of cathode and grid wire conditioning, improvements to the krypton calibration system, as well as the xenon controls and recovery system. A D-D neutron generator providing a near monochromatic source of neutrons was used to make an \emph{in situ} calibration (down to 0.7 keV for the ionization channel) of the low-energy nuclear recoil response of LUX through an analysis of multiple-scatter events~\cite{Verbus:2014}. The sensitivity for the 300-day run is expected to surpass that of the first WIMP-search result by a factor of around five and the sensitivity at low masses will benefit from the confirmation of the detector response to low-energy recoils.

\section{LUX-ZEPLIN} 

\begin{figure}[!ht]
\begin{center}
\includegraphics[width=0.8\columnwidth]{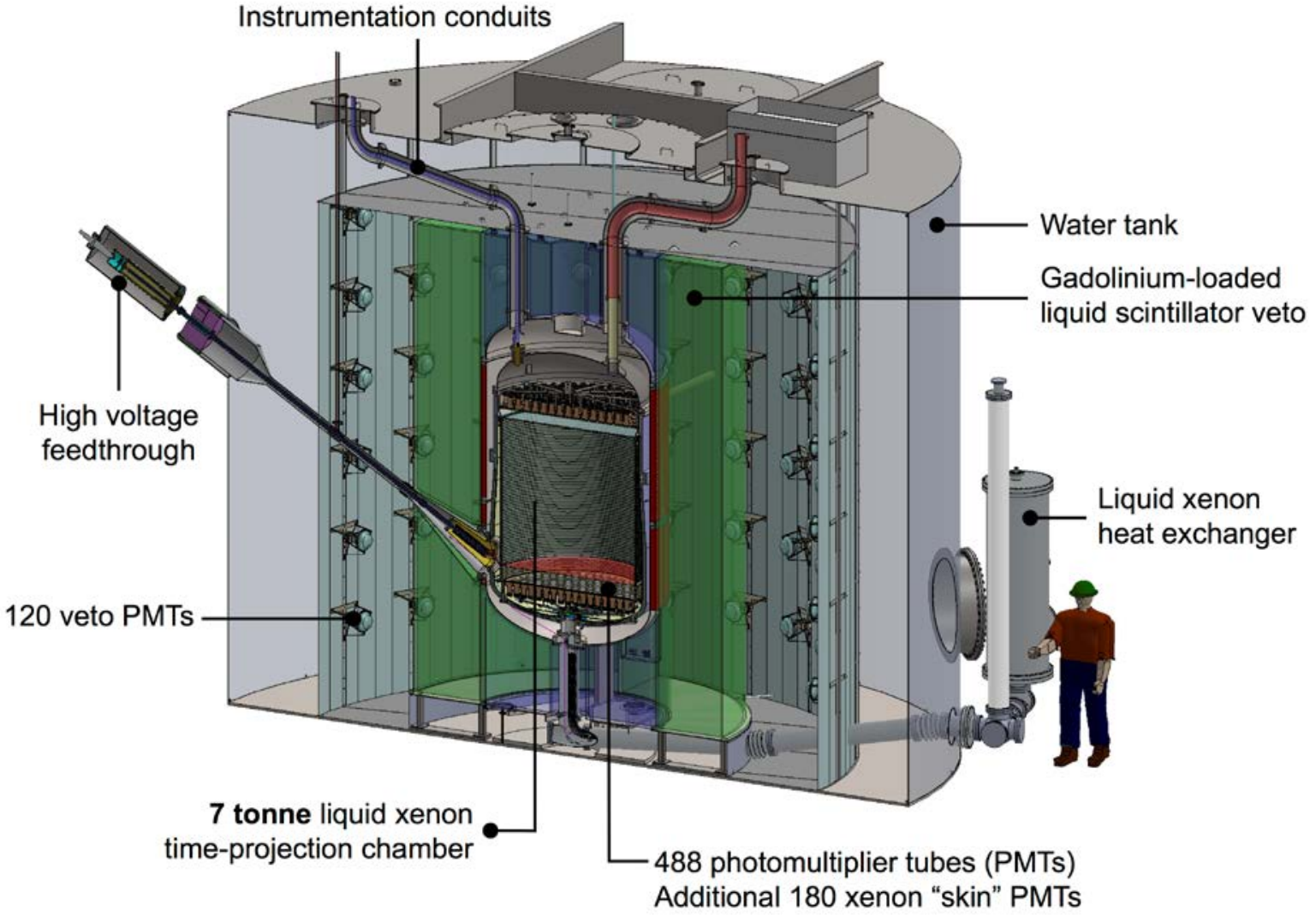}
\caption{Schematic of the LZ experiment as housed in the reused LUX water-tank.}
\label{fig:lz}
\end{center}
\end{figure}

The LUX-ZEPLIN (LZ) experiment will operate a scaled up version of the LUX TPC with an active region containing about 7 tonnes of liquid xenon. LZ will replace LUX on the 4850' level at SURF and will reuse the LUX water tank. Figure~\ref{fig:lz} shows the overall detector concept. In addition to the considerable increase in target mass ($\sim$45 $\times$ LUX fiducial) LZ will have lower background construction and will features a high efficiency veto system including an active xenon {\em skin} layer between the TPC and the cryostat, and an external liquid scintillator veto (gadolinium loaded linear alkyl benzene). The combination of skin readout and the outer detector provides powerful rejection of $\gamma$-rays and neutrons from internal sources (e.g. PMTs), and confidence in interpretation of possible signal. 

With a projected sensitivity of $10^{-48} \mathrm{cm}^{2}$ for its full 1000-day exposure, LZ will have sensitivity unmatched by any competing experiment on a similar timescale, exploring a significant fraction of the unexplored electroweak parameter space remaining above the irreducible background from coherent scattering of neutrinos from astrophysical sources~\cite{Billard:2013qya}. 

\section{Summary}
	
LUX has set set the world's most stringent limit for spin-independent WIMP-nucleon elastic scattering, becoming the first direct search experiment to probe the sub-zeptobarn regime. The LUX 300-day run will further increase this sensitivity by a factor of five with discovery still possible. Beyond LUX, the LUX-ZEPLIN experiment will improve on LUX by almost two orders of magnitude in WIMP-nucleon interaction sensitivity, enabling significantly deeper probing of parameter space for discovery if necessary, or giving the capability to characterise a dark matter signal if found.

\bigskip
\section{Acknowledgments}

his work was partially supported by the U.S. Department of Energy (DOE) under award numbers DE-FG02-08ER41549, DE-FG02-91ER40688, DE-FG02-95ER40917, DE-FG02-91ER40674, DE-NA0000979, DE-FG02-11ER41738, DE-SC0006605, DE-AC02-05CH11231, DE-AC52-07NA27344, and DE-FG01-91ER40618; the U.S. National Science Foundation under award numbers PHYS-0750671, PHY-0801536, PHY-1004661, PHY-1102470, PHY-1003660, PHY-1312561, PHY-1347449; the Research Corporation grant RA0350; the Center for Ultra-low Background Experiments in the Dakotas (CUBED); and the South Dakota School of Mines and Technology (SDSMT). LIP-Coimbra acknowledges funding from Funda\c{c}\~{a}o para a Ci\^{e}ncia e Tecnologia (FCT) through the project-grant CERN/FP/123610/2011. Imperial College and Brown University thank the UK Royal Society for travel funds under the International Exchange Scheme (IE120804). The UK groups acknowledge institutional support from Imperial College London, University College London and Edinburgh University, and from the Science \& Technology Facilities Council for PhD studentship ST/K502042/1 (AB). The University of Edinburgh is a charitable body, registered in Scotland, with registration number SC005336.

%
%

%
%
%
%
 
\end{document}

%% file: econfmacros.tex
\definecolor{Red}{rgb}{1,0,0}
\definecolor{Green}{rgb}{0,1,0}
\definecolor{Blue}{rgb}{0,0,1}
\definecolor{Black}{rgb}{0,0,0}

\def\beq{\begin{equation}}
\def\eeq#1{\label{#1}\end{equation}}
\def\eeqn{\end{equation}}

\def\beqa{\begin{eqnarray}}
\def\eeqa#1{\label{#1}\end{eqnarray}}
\def\eeqan{\end{eqnarray}}

\let\bar=\overbar

\def\Dslash{\not{\hbox{\kern-4pt $D$}}}
\def\dslash{\not{\hbox{\kern-2pt $\del$}}}

\def\msb{{\bar{\ssstyle M \kern -1pt S}}}